\documentclass{PoS}

\usepackage{amsmath}

\usepackage{siunitx}
\newcommand{\lrp}[1]{\left(#1\right)}
\newcommand{\lra}[1]{\left< #1 \right>}

\title{Search for very-high-energy gamma-ray counterparts of gravitational waves with HAWC}

\ShortTitle{Search for counterparts of gravitational waves with HAWC}

\author{\speaker{Israel Martinez-Castellanos}\\
        University of Maryland\\
        E-mail: \email{imc@umd.edu}}

\author{For the HAWC Collaboration\\
        For a complete author list and acknowledgement see PoS(ICRC2019)1177.
        \href{https://www.hawc-observatory.org/collaboration/icrc2019.php}{https://www.hawc-observatory.org/collaboration/icrc2019.php}}

\abstract{The High-Altitude Water Cherenkov Observatory (HAWC) is a large field of view (\SI{\sim2}{sr}) continuously operating experiment sensitive to very-high energy (VHE) gamma rays (\SIrange{\sim0.3}{100}{TeV}). These characteristics make it well suited for observing or constraining the VHE emission of rapid transients such as some gravitational waves progenitors. Of special interest are the events at low redshift where the attenuation due to the extragalactic background light is minimal. This is the case for binary neutron star mergers in the horizon of the LIGO and Virgo experiments, for which HAWC can either detect or place constraining limits on events occurring in our field of view. We report on our search for counterparts of the gravitational waves detected by LIGO and Virgo.}

\FullConference{36th International Cosmic Ray Conference -ICRC2019-\\
		July 24th - August 1st, 2019\\
		Madison, WI, U.S.A.}

\begin{document}

\section{Introduction}

Gravitational waves (GWs) are disturbances in the metric of spacetime. They have long been predicted to exist since they are a solution of Einstein's field equations. They manifest as a change of proper distance in the plane transverse to the direction of propagation. Current gravitational wave detector consist on highly sensitive kilometers-long interferometers that measure relative arm length variations by a fraction of \num{\sim e-20} or less \cite{aLIGO-detector-2015}. 

Gravitational waves are expected from any source with a varying quadrupole moment. Compact binary objects in orbit, such as neutron stars and black holes, are one of the primary sources of gravitational waves. The Laser Interferometer Gravitational-Wave Observatory (LIGO) successfully detected in 2015 the inspiral of a binary black hole (BBH) \cite{GW150914-ligo}.

Although the merger of two black holes is not expected to result in an electromagnetic counterpart (unless, maybe, under some special circumstances \cite{charged-bbh-Zhang-2016,grb-bbh-Loeb-2016}), the merger of a binary neutron star (BNS) or a neutron start-black hole binary (NSBH) are believed to be the progenitors of short duration gamma-ray bursts (GRBs) \cite{sgrb-berger-review-2014,fong-sgrb-review-2015}. The detection of GW170817/GRB170817a \cite{GW170817-multimessenger}, along with the measurements across the electromagnetic spectrum, is strong evidence this is the case, at least for some short GRBs. Note that supernovae, believed to be the progenitors of long duration GRBs \cite{grb-supernova-hjorth2012,longgrb-corecollapse-diffenv-Fruchter2006}, can also result in the emission of gravitational waves due to non-symmetric mass distributions.

The joint observation of gravitational waves and electromagnetic emission from an event is very fruitful. The former can provide information about the distance, masses, nature of the progenitor, relative viewing angle, time of coalescence, etc. \cite{LALinference-2015}; the latter can provide, for example, information about the environment, jet geometry, redshift,  and total energy emitted. The VHE observations can probe properties such as the jet bulk Lorentz factor and test models about the the interactions with the surrounding medium and emission mechanisms \cite{review-grb-physics-kumar-2015, grb-tev-totani-1998,constraining-tev-grb-Fragile-2004}. 

Observing the VHE emission of GRBs is challenging, partly because it can be heavily attenuated by the Extragalactic Background Light (EBL). However, GRBs occurring within the detection range of current gravitational wave detectors (at most a few hundreds of megaparsecs at design sensitivity \cite{prospects-gw-2018}, or $z \lesssim 0.1$) are not expected to be severely attenuated by EBL \cite{ebl-gilmore-2012}. In any case, in order to study this during the prompt phase, where the bulk of the high-energy emission is observed, an instrument with a wide field of view is required. 

The High-Altitude Water Cherenkov Observatory (HAWC) is a surface array containing 300 detector units over an area of \SI{22000}{m^2} (see \cite{crabPaper-Abeysekara-2017} for details). It is sensitive to gamma rays between \SI{\sim 100}{GeV} to \SI{\sim 100}{TeV}, and operates with a high duty cycle (\num{>95}\%) continuously monitoring the overhead sky with an instantaneous field of view of \SI{\sim 2}{sr}. These characteristics make it well suited to study transient phenomena. In this work we use it to search for short timescale emission correlated with the gravitational waves detected by LIGO and Virgo during the observation runs O1 (from 2015-09-12 to 2016-01-19), O2 (from 2016-11-30 to 2017-08-25) and the ongoing O3 (started on 2019-04-01, expected to last one calendar year). 

\section{Search method}

In this work we search for a significant excess in short timescales against the cosmic ray background. The background is estimated by convolving the local direction arrival probability, which is mostly constant, with the total event detection rate, which can fluctuate by \num{\sim10}\%. This method is called \textit{direct integration}\cite{direct_integration_Atkins_2003} and can estimate the background up to 1 part in \num{e3}, the relative intensity of the cosmic ray anisotropy.   

In order to test the precense of a source against the background-only hypothesis a binned maximum likehood is used. This is implemented as described in \cite{crabPaper-Abeysekara-2017}, but adapted to handle short timescales with an analysis framework internally called ZEnith Band Response Analysis (ZEBRA). The data is divided into 9 analysis bins based on the fraction of channels that detected a signal, $f_{hit}$. This is correlated with energy, increasing as a function of $f_{hit}$. It is also correlated with the quality of the event: the higher the $f_{hit}$ the better the angular resolution --ranging from \ang{\sim 0.1} to \ang{\sim 1}-- and the cosmic-ray background suppression. Using a binned likelihood allows us to make full use of the increase in sensitivity versus energy. For events within the LIGO and Virgo range, where the attenuation of high-energy events due to EBL is minimal, this results in a significant increase in sensitivity with respect to a classical ON/OFF analysis used in previous HAWC GRBs searches \cite{dirk-grb-2017}. 

For a given sky location and time window we assume a source with a simple power law spectrum
\begin{align}
\phi(E) = \phi_0 \lrp{\frac{E}{E_0}}^{\alpha} \,,
\end{align}
where $\phi_0$ is the normalization and $\alpha$ the spectral index. This hypothetical spectrum is modified by the expected EBL attenuation based on \cite{ebl-gilmore-2012} and the most probable distance according to gravitational wave data. The sensitivity of the search is only weakly dependent on the assumed index, which we set to $\alpha = -2$. We then compare the likelihood ratio between the source (with a free normalization) and background-only hypotheses (setting the normalization $\phi = 0$), and use it as a test statistic
\begin{align}
TS = \max\lrp{ 2 \log \frac{\mathcal{L}\lrp{\phi_0}}{\mathcal{L}\lrp{\phi_0 = 0}}} \,.
\end{align}

We search in the region containing 95\% of the gravitational wave sky localization probability. We limit the search to locations within \ang{45} from zenith since the sensitivity rapidly decreases for larger zenith angles. The search is performed in a HEALPix grid \cite{HEALPix_Gorski_2005} with a spacing of \ang{0.11} ($N_{side} = 512$), selected based on our angular resolution.

Eight different time window widths are tested, $\Delta t = $~\SI{0.3}s, \SI{1}s, \SI{3}s, \SI{10}s, \SI{30}s, \SI{100}s, \SI{300}s and \SI{1000}s. This is motivated by the typical emission duration of short GRBs detected by Fermi-LAT \cite{second-lat-grb-catalog}, but also allows for a possible difference between the high-energy and very-high-energy prompt emission timescale. The time windows are shifted by $0.2\Delta t$ starting at $t_0-5\Delta t$ and stopping at $t_0+10\Delta t$, where $t_0$ corresponds to the time of coalescence of the binary merger. This assumes the emission duration is of a similar order of magnitude as the delay with respect to the coalescence. While we expect gamma-ray emission to occur after coalescence, we analyze also a period of time previous to the GW trigger to allow for possible unexpected phenomena. 

For a given time window width we estimate the $TS$ distribution under the null hypothesis by randomly fluctuating the expected background and running the same analysis we do on data. This accounts for the correlation between nearby pixels in the grid and neighboring time windows caused by oversampling. Using this distribution we obtain the false alarm ratio per solid angle (FAR) for the results of each tested time window width. The simulated results from each timescale are conservatively considered independent, as the correlations would be computationally prohibited to estimate otherwise. We then estimate the expected number of events with $TS$ greater than a given obtained result as
\begin{align}
\lra{n}_{\Delta t} = \text{FAR}_{\delta t} \times \Delta T_{\Delta t} \times A \times N
\end{align} 
where $A$ is the total solid angle searched, $\Delta T_{\Delta t}$ is the total time searched (typically $15 \Delta t$) and $N$ is the total number of timescales tested ($N=8$ in this work). The lowest value of $\lra{n}_{\Delta t}$ is considered the $p$-value of the search, which is valid when $\lra{n} \ll 1$.  

Starting O3, this search has been running automatically without human intervention after each gravitational wave alert is received. The latency goes from a few minutes to \SI{1}{hr}, depending of the area covered by the sky localization probability. In case of detection, HAWC will be able to constrain the localization uncertainty to $\lesssim$\ang{0.5} and help other telescopes to observe it.

\section{Sensitivity and expectations}

\begin{figure}
\centering
\includegraphics[width=.8\textwidth]{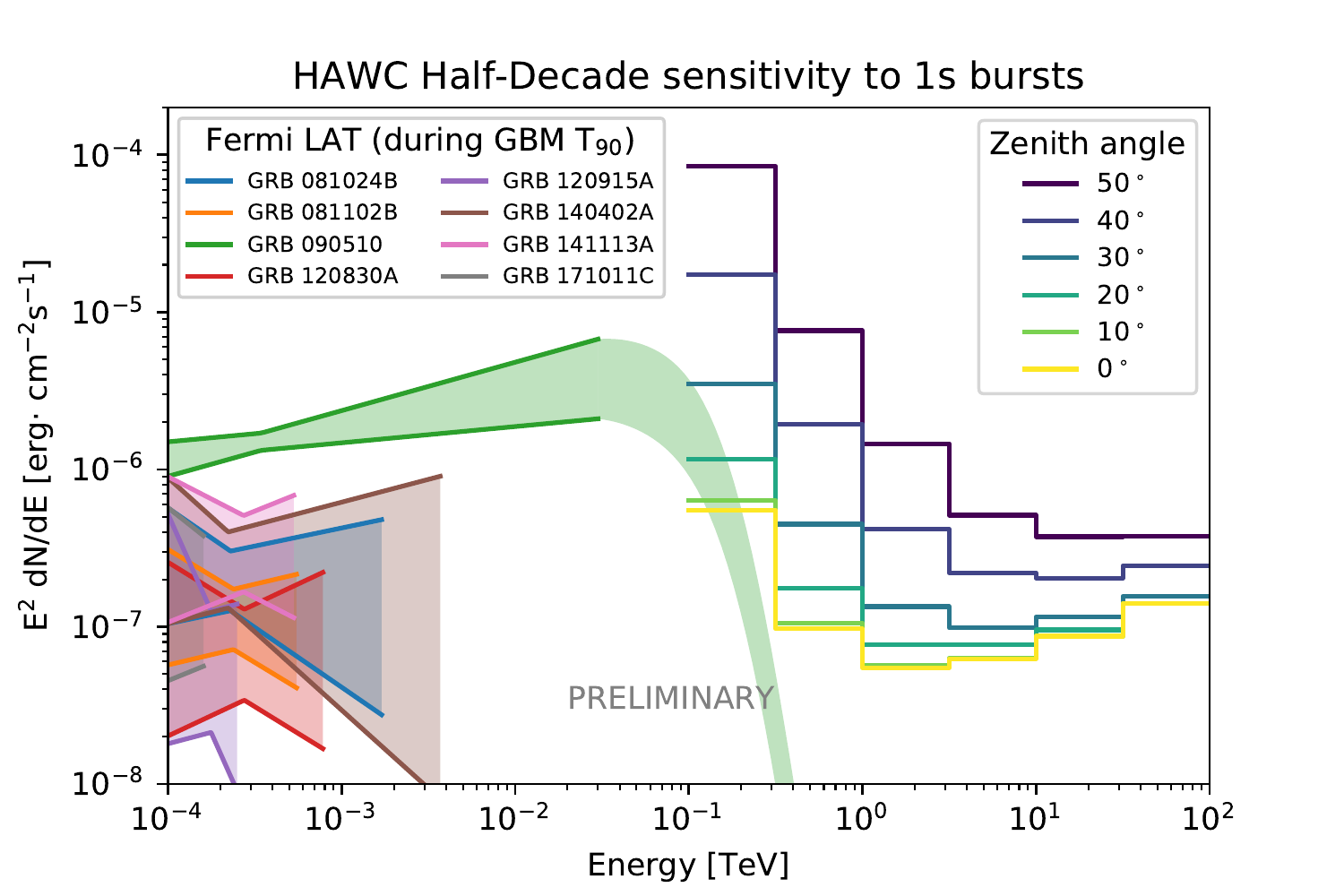}
\caption{HAWC quasi-differential sensitivity to 1s bursts as a function of the zenith angle, defined as the mean flux in a given half-decade that would result in at least a 5$\sigma$ detection half of the time. For reference we include the spectrum for the short GRBs detected by the Fermi-LAT as measured in the Fermi-GBM $T_{90}$ window, which is of the order of 1s, reported in \cite{second-lat-grb-catalog}. The redshift is known only for GRB 090510 ($z=0.9$); its spectrum was extrapolated using a simple power law and attenuated by EBL.}
\label{fig:grb-sensi-1s}
\end{figure}

\begin{figure}
\centering
\includegraphics[width=.8\textwidth]{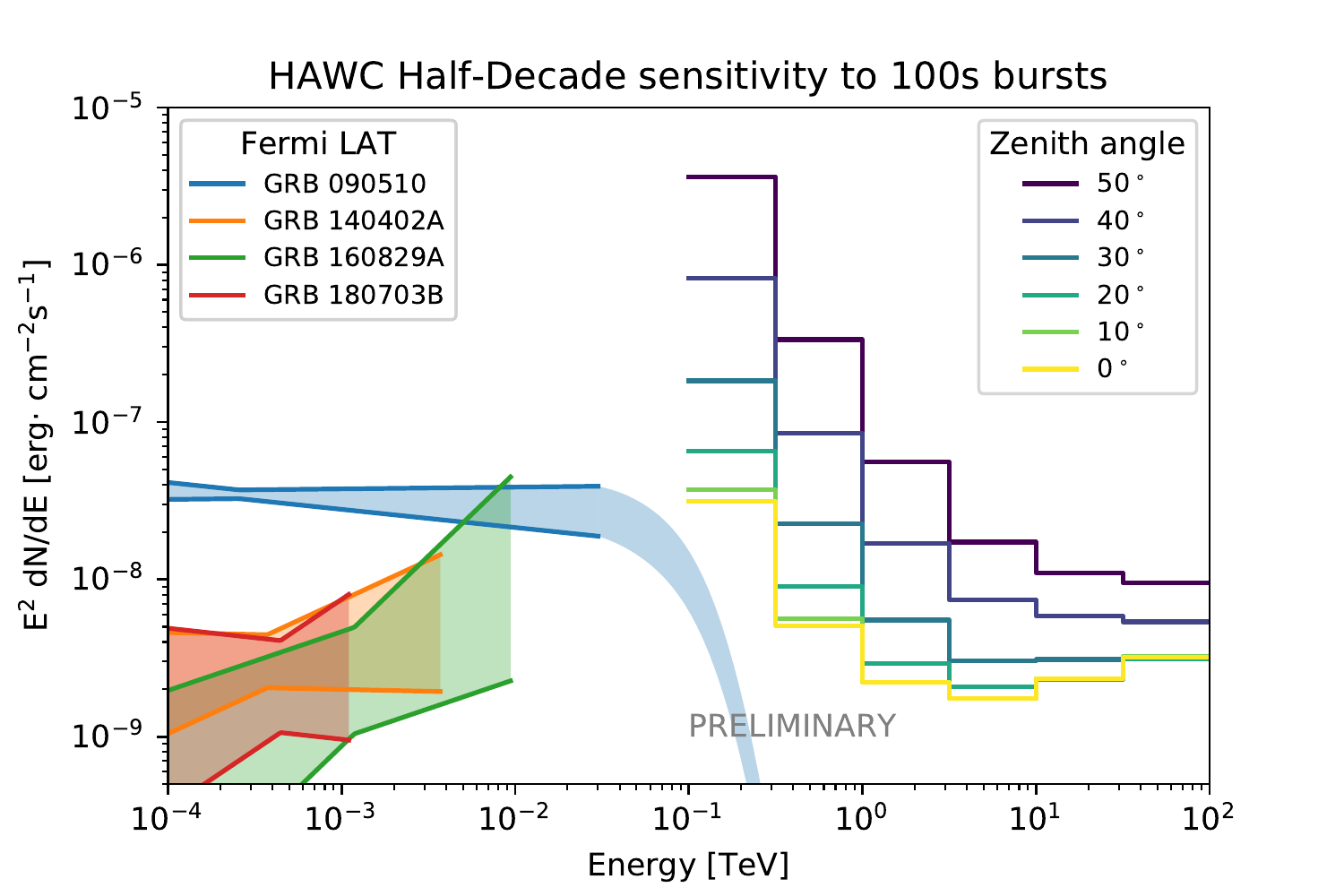}
\caption{HAWC quasi-differential sensitivity to 100s bursts. We show for reference the spectra measured by the Fermi-LAT for GRBs with a duration \SI{>70}{s} (the longest being \SI{170}{s}). See description of Figure \ref{fig:grb-sensi-1s} for more details.}
\label{fig:grb-sensi-100s}
\end{figure}

Due to the lack of observations, it is not clear what the characteristics of VHE gamma-ray counterparts of gravitational waves would be. A probable scenario, such as for GW170817, is the presence of a short GRB accompanying at least some gravitational waves originated by the merger of a binary neutron star. Neutron star-black hole mergers might also result in short GRBs, although their merger rate is expected to be lower \cite{merger-rate-ligo-2010}. The characteristics of the VHE prompt emission of a short GRB is also unknown, but one can speculate it might have a similar timescale as the HE emission detected by Fermi-LAT. For the 17 short GRBs detected, the time of the emission ranges from a fraction of a second to \SI{\sim 170}{s} \cite{second-lat-grb-catalog}. Motivated by this, in Figures \ref{fig:grb-sensi-1s} and \ref{fig:grb-sensi-100s}  we show the HAWC differential sensitivity to \SI{1}{s} and \SI{100}{s} burst, respectively.

The \SI{100}{s} burst sensitivity is compared to GRBs with an emission in the LAT lasting \SI{>70}{s}. For the \SI{1}{s} burst sensitivity, instead of comparing to the average spectrum, we choose the measurement (by the LAT) during the period containing 90\% of the emission detected by the GBM, which is \SI{\sim 1}{s} long.  We chose this reported value instead of the average since the time window over which the LAT detected emission is not necessarily optimal for HAWC. Although the number of passing events is low for these timescales, HAWC is still background dominated and is more sensitivity to the peak flux rather than the total fluence. Note that since the emission detected by Fermi-LAT is typically delayed compared to the one measured by Fermi-GBM, this is not expected to be the optimal \SI{1}{s} window. Roughly half of the short GRBs do not have a measured emission by the LAT during this period.

It is however not straightforward to extrapolate the spectrum measured by Fermi-LAT to the VHE range. On one hand, although in general the emission in this energy range can be heavily attenuated by EBL, GRBs occurring within the range of current gravitational wave detector would be close enough that this effect would be minimal. On the other, significant attenuation could also occur at the source due to photon-photon interactions. If there is no attenuation and the spectrum continues up to \SI{\sim 1}{TeV}, it is plausible that HAWC could detect short GRBs such as the ones in Figure \ref{fig:grb-sensi-1s}. Furthermore, it is reasonable to expect nearby GRBs within the LIGO-Virgo range to be significantly brighter than the average, improving the chance of detection. This might not hold true, as shown by GRB 170817A for which Fermi-GBM recorded an ordinary flux \cite{fermi-gbm-grb170817} in spite of being the closest GRB with a measured distance to date.

Despite all the unknowns and difficulties predicting the possible VHE gamma-ray counterpart that might accompany some gravitational waves, HAWC is in a position to either detect it or set a meaningful constraint if the event is known to have occurred in our field of view. 

\section{Results}

HAWC was taking data in normal operations during all gravitational wave events detected by LIGO and Virgo during O1, O2 and O3 (analyzed here until June 29, 2019). We analyzed all the events with at least a partial overlap with the HAWC field of view. Note that this is completely determined by the date and time based on the HAWC location (\ang{19;1;49.78} \SI{1}{\arcmin}  N, \ang{97;16;11.454} W). In particular, we analyzed 25\%, 28\% and 4\% of the sky localization probability of S190425z, S190426c and S190510g, events whose progenitors are likely to contain a neutron star. The location of GW170817 was not in the HAWC field of view at the time of the merger. 

The results for all events are consistent with background expectations. The maximum test statistics at any location, time, event or timescale was $TS=26.0$, corresponding to a $p$-value of 1.0 after trials. 

\begin{figure}
\centering
\includegraphics[width=.6\textwidth]{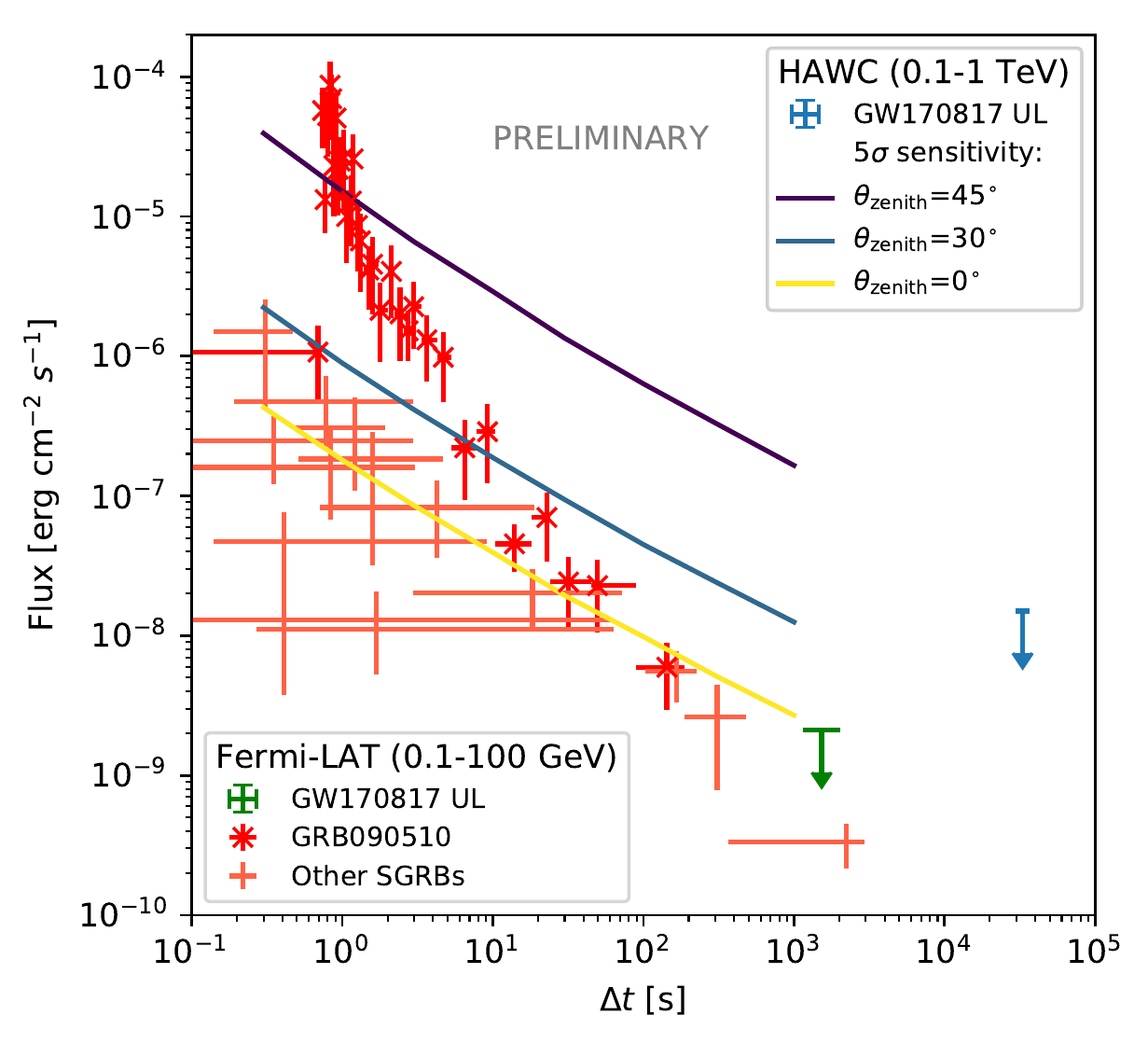}
\caption{HAWC \SI{95}{\percent} C.L. upper bound (blue) on GW170817 between \SI{100}{GeV} and \SI{1}{TeV}  from $\Delta t = t - t_0 =  $~\SI{8.20}{hrs} to $\Delta t = $~\SI{10.23}{hrs} where $t_0$ is the time of coalescence. Fermi-LAT UL between \SI{100}{MeV} and \SI{100}{GeV}  from  $\Delta t = $\SI{1153}{s} to $\Delta t = $\SI{2027}{s} are shown in green. For comparison we show the HAWC sensitivity (\SIrange{0.1}{1}{TeV}) a function of the emission duration $\Delta t$ and Fermi-LAT measurements of previously detected short GRBs (red) \cite{fermi-lat-gw170817-2017}. The Fermi-LAT measurements of GRB090510 as a function of time are highlighted.}
\label{fig:GW170817UL}
\end{figure}

Since AT 2017gfo, the optical counterpart of GW170817, was not in our field of view at the time of the gravitational wave trigger, we analyze the emission during the following transit, from \SI{8.20}{hrs} to \SI{10.23}{hrs} after. No significant emission was found. We set an upper limit of \SI{1.5e-8}{ergs.cm^{-2}.s^{-1}} between \SI{100}{GeV} and \SI{1}{TeV} during this period, shown in Figure \ref{fig:GW170817UL}. This event transited at the edge of our field of view, which resulted in a degraded sensitivity. 

\section{Conclusions}

We have searched in HAWC data for the potential very-high-energy gamma-ray counterparts of gravitational waves detected by LIGO and Virgo during the observation runs O1, O2 and O3. No significant emission was observed. However, we showed that our sensitivity compared to short GRBs detected at high-energy indicates that HAWC should be able to either detect or set constraining upper limits for events occurring in our field of view and within the range of current gravitational wave detectors. 

As a wide-field-of-view instrument, HAWC is in a unique position to study the prompt emission of an associated gamma-ray burst. Furthermore, it can be instrumental to constrain the event localization, helping other telescopes to study the electromagnetic counterpart. With this goal, starting O3, HAWC follows-up automatically gravitational wave alerts allowing a low response latency. We will continue to perform this search and inform the community of our results.

\bibliographystyle{unsrt}
\bibliography{Bibliography}

\end{document}